# An Aboriginal Australian Record of the Great Eruption of Eta Carinae

**Duane W. Hamacher**
Department of Indigenous Studies, Macquarie University, NSW, 2109, Australia
duane.hamacher@mq.edu.au

**David J. Frew**
Department of Physics & Astronomy, Macquarie University, NSW, 2109, Australia
david.frew@mq.edu.au

## Abstract

We present evidence that the Boorong Aboriginal people of northwestern Victoria observed the Great Eruption of Eta (η) Carinae in the nineteenth century and incorporated the event into their oral traditions. We identify this star, as well as others not specifically identified by name, using descriptive material presented in the 1858 paper by William Edward Stanbridge in conjunction with early southern star catalogues. This identification of a transient astronomical event supports the assertion that Aboriginal oral traditions are dynamic and evolving, and not static. This is the only definitive indigenous record of η Carinae's outburst identified in the literature to date.

**Keywords:** Historical Astronomy, Ethnoastronomy, Aboriginal Australians, stars: individual (η Carinae).

## 1    Introduction

Aboriginal Australians had a significant understanding of the night sky (Norris & Hamacher, 2009) and frequently incorporated celestial objects and transient celestial phenomena into their oral traditions, including the sun, moon, stars, planets, the Milky Way and Magellanic Clouds, eclipses, comets, meteors, and impact events. While Australia is home to hundreds of Aboriginal groups, each with a distinct language and culture, few of these groups have been studied in depth for their traditional knowledge of the night sky. We refer the interested reader to the following reviews on Australian Aboriginal astronomy: Cairns & Harney (2003), Clarke (1997; 2007/2008), Fredrick (2008), Haynes (1992; 2000), Haynes et al. (1996), Johnson (1998), Norris & Hamacher (2009), and Tindale (2005).

The first detailed publication on Aboriginal astronomy in the literature was by William Stanbridge, who described the astronomy and mythology of the Boorong clan of the Wergaia language from the dry Mallee country near Lake Tyrell in northwest Victoria (Stanbridge, 1858; 1861, see Figure 1). The Boorong word for Tyrell (tyrille) meant "sky" and they prided themselves on knowing more astronomy than any other Aboriginal community (Stanbridge, 1858:137; 1861:301). Stanbridge read his seminal paper to the Philosophical Institute of Victoria on 30 September 1857. He wrote:

> "*I beg to lay before your honorable Institute the accompanying paper on the Astronomy and Mythology of the Aborigines, and in doing so I am sensitive of its imperfectness, but as it is now six years since I made any additions to it, and as my occupation does not lead me to that part of the country where I should be able to make further additions, I have presumed to present it to your society, hoping that it may be a means of assisting with others to gather further traces of the people that are so fast passing away.*
>
> *This statement of the Astronomy and Mythology of the Aborigines is, as nearly as language will allow, word for word as they have repeatedly during some years stated it to me. It is in the language of, and has been gleaned from, the Booroung Tribe, who claim and inhabit the Mallee country in the neighbourhood of Lake Tyrill, and who pride themselves upon knowing more of Astronomy than any other tribe.*" (Stanbridge, 1858:137).

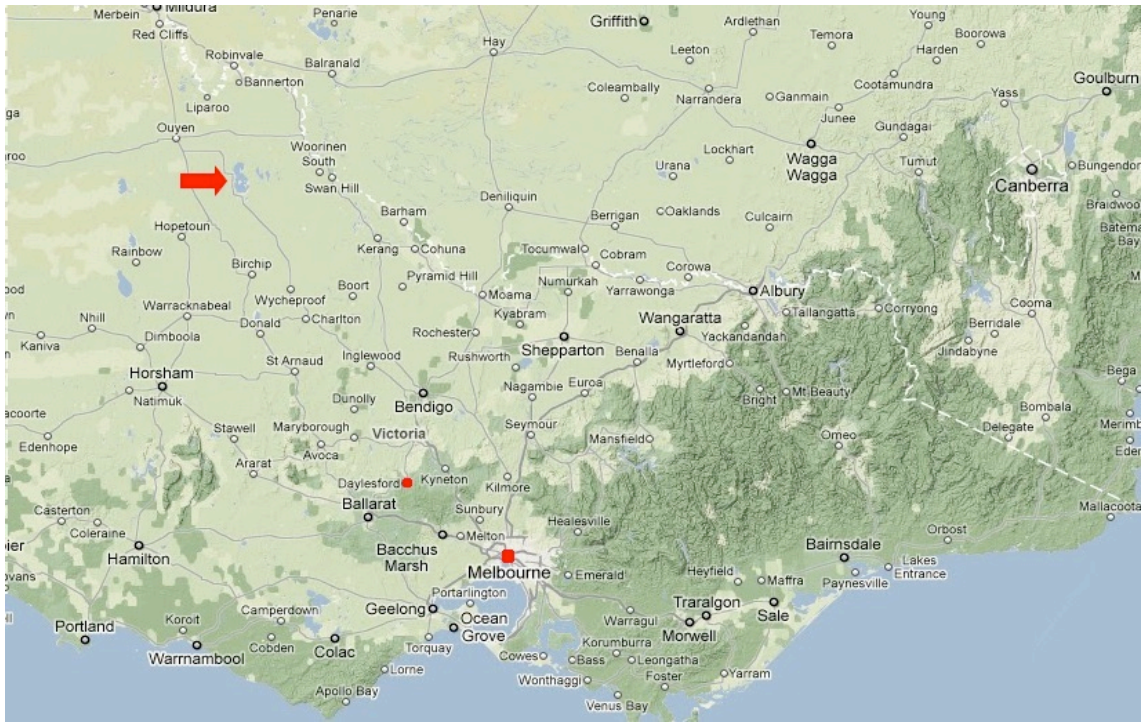

**Figure 1:** A map of Victoria (south-eastern Australia). Lake Tyrell, where Stanbridge acquired his knowledge of Boorong astronomy, is indicated by the red arrow, and Tyrell Downs is just to the east of Lake Tyrell. Melbourne, where he lectured to the Philosophical Institute of Victoria, and Daylesford, were he lived for a time, are marked by red dots

Stanbridge's work describes Boorong views of various celestial objects and phenomena, including the Sun, Moon, Jupiter, Venus, numerous individual stars, the Pleiades and Coma Berenices open star clusters, two compact constellations (Delphinus and Corona Australis), the Magellanic Clouds, the Coalsack Nebula, the Milky Way, meteors, and the seasons. These celestial objects are represented by characters in oral traditions that are typically represented by animals or beings, and occasionally their spouses. In some cases, Stanbridge identified the star by name, while in other cases it is identified only by a general description, which typically includes its brightness, constellation, and proximity to particular stars. Stanbridge's work was later re-analysed by MacPherson (1881) and Morieson (1996).

In this paper, we identify one of Stanbridge's Boorong descriptions as referring to the outburst of the super-massive binary system Eta (η) Carinae. We begin by providing a brief biographical account of William Stanbridge in Section 2, while in Section 3 we independently identify all the celestial objects described in Stanbridge (1858). In Section 4 we summarise the evidence that indicates that the Boorong observed η Carinae during its Great Eruption in the nineteenth century and incorporated it into their oral traditions, while we discuss η Carinae in Section 5. In Section 6, we discuss previous attempts to identify the wife of War (the crow), while in Section 7 we use this as an exemplar to show that the Boorong incorporated η Carinae into their oral traditions during the nineteenth century period of outburst and not before. In Section 8 we explain how transient astronomical phenomena are often incorporated into oral traditions and show that sky knowledge is dynamic and changing, while in Section 9 we search for other indigenous records of η Carinae, before summarising our conclusions in Section 10.

## 2      William Stanbridge: biographical details

William Edward Stanbridge (Esq, M.L.C., J.P., see Figure 2) was born on 1 December 1816 in the village of Astley in Warwickshire, England to Edward and Anne Stanbridge (Parliament of Victoria, 2010). In November 1841, at the age of 24, Stanbridge arrived in Port Phillip, Victoria and moved around Victoria and South Australia over the next several years, finally settling near Daylesford, Victoria (Billis & Kenyon, 1974:143). Soon after his arrival in Daylesford (1851-1852), Stanbridge purchased the Holcombe run and later Wombat run, both north of Daylesford (The Argus, 1848:1; Billis & Kenyon, 1932:221). He was issued a pastoral license for Tyrrell Station (The Argus, 1848:1), from September 1847 to January 1873, and was the first non-indigenous person to do so (Billis & Kenyon, 1932:143).

On 8 July 1862, Stanbridge was appointed "honorary correspondent for the Upper Loddon district, of the control board for watching over the interests of the aborigines" (The Argus, 1862:5). As a pastoralist and investor, he became wealthy from gold mining and later married Florence Colles on 21 August 1872 (The Argus, 1872:4), who died during the birth of their daughter, Florence Colles Stanbridge, on 1 August 1878 (The Argus, 1878:1). Later that year, as a memorial to his late wife, he founded the Florence Stanbridge Scholarship at Trinity College, Melbourne, where he was member of the College Council (The Argus, 1881:6). Stanbridge became a member of the Philosophical Institute of Victoria from 1857-1859 (The Argus, 1857:5), the Royal Society of Victoria in 1860 (Royal Society of Victoria, 2010), was a Fellow of the Anthropological Institute, London, and a member of the Church of England Assembly (Parliament of Victoria, 2010). He became the Member First Council and first chairman of Daylesford, later the Councilor (1868-1874, 1880-1892) and finally the mayor (1882-1883; Thomson & Serle, 1972). Upon his death on 5 April 1894 in Daylesford, aged 77, Stanbridge left funds in his will to found the Frances Colles Stanbridge Scholarship (named after his mother-in-law) at Trinity College (The Argus, 1895:6).

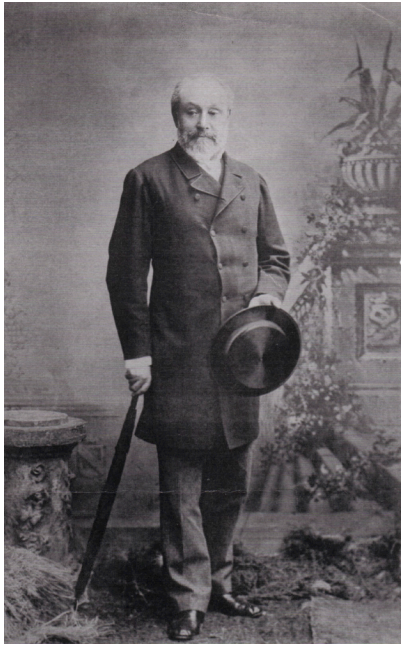

**Figure 2:** One of the few extant photos of William Edward Stanbridge, courtesy of Keva Lloyd. The date of the photograph is unknown, but is probably around ca. 1880s.

Stanbridge presented his paper on Boorong astronomy to the Philosophical Institute of Victoria in Melbourne on 30 September 1857 (which was published in their 1858 proceedings). A second, longer paper was later published in 1861 and included all the essential astronomical information from the original 1858 manuscript, which was apparently destroyed in a fire (Stanbridge, 1861:304). Stanbridge claimed to have gained his information on Boorong astronomy from two members of a Boorong family who had the reputation of having the best astronomical knowledge in the community (Stanbridge, 1858:304). He stated that his first fieldwork experience was conducted by a small campfire under the stars, located on a large plain near Lake Tyrell (Stanbridge, 1861:301,304). In his original 1857 address, Stanbridge said that he had not made any additions to the paper in the previous six years, implying that he did the bulk of his astronomical fieldwork between being issued a pastoral license at Tyrell Downs (ca. 1848) and 1851, with the latter date being better constrained.

His education and training in astronomy is unclear, but his written papers show that he had an acceptable knowledge of astronomy and navigation. It is also uncertain what astronomy references he had at his disposal while conducting his fieldwork, but the most widespread contemporary star catalogue was the "British Association Catalogue" (Baily, 1845). Other possibilities include the Parramatta or 'Brisbane' catalogue (Richardson, 1835), the complete edition of La Caille's catalogue of 9,766 stars (Henderson & Baily, 1847), or the earlier La Caille (1763) catalogue of 1,942 southern stars, which will be discussed further in Section 4.

## 3        Identification of Boorong Celestial Objects

Stanbridge (1858, 1861) provides equivalent western names for the majority of stars identified by the Boorong, which were repeated in Smyth (1876:432-434). As part of our re-analysis of Stanbridge's work, we seek to independently identify these stars, as well as all the stars he did not specifically name. Since the Boorong clan apparently no longer exists as an entity and much of their traditional knowledge has been lost (Morieson, 1996), Stanbridge's papers are the only primary source of Boorong ethnoastronomy. We use Stanbridge (1858) to identify all stellar objects that are not identified by name, which we present in Table 1. We then provide a complete list of all the Boorong celestial objects and seasons identified by Stanbridge and ourselves in Table 2, including visual magnitudes for all stars.

Some of our proposed object identifications need additional clarification. The Boorong object called Totyarguil is the bright star Altair rather than the whole constellation of Aquila, since Stanbridge (1858:139) specifically noted that "the stars on either side are his two wives". This comment certainly refers to β Aql and γ Aql, which flank Altair prominently in the sky; MacPherson (1881:76) came to a similar conclusion. Similarly, Neilloan (a flying Loan or Mallee Fowl) is identified in Stanbridge with an object called Lyra. Stanbridge's description does not suggest that Neilloan refers to a group of stars. We note that prior to the late nineteenth century, Vega was commonly called 'Lyra' or 'Lucida Lyrae' in the literature (e.g. Herschel, 1847:334), so we match Neilloan to the bright, zero-magnitude star Vega, as distinct from the whole constellation of Lyra, as instead proposed by Morieson (1996)[1]. In agreement with our conclusion, MacPherson (1881) also identified Neilloan to be the star Vega.

We also note the explicit mention of the relatively inconspicuous star Sigma (σ) Canis Majoris (CMa) by Stanbridge. Perhaps it was pointed out to Stanbridge because it lies between the bright stars Wezen (δ CMa) and Adhara (ε CMa) on an approximate straight line (an apparent positional preference of the Boorong noted by MacPherson, discussed below). Alternatively, and less likely, it may have been noticed as being variable in brightness. In fact, this reddish star ($V = +3.45$, B-V = 1.72; spectral class M0-Ib), is a known irregular variable with a small amplitude of ~ 0.1 mag (Samus et al., 2010). There is a suggestion of larger amplitude variability in the past (J.E. Gore, quoted by Chambers, 1875), but this has not been confirmed by modern observations. Even so, there is a small possibility that the star may have been substantially brighter in the nineteenth century[2].

As shown in Table 2, the Boorong created a reasonably good list of the brightest stars visible from northwestern Victoria. The only first magnitude stars omitted by Stanbridge were (with the $V_{mag}$ in parentheses): Procyon (+0.40), Betelgeuse (+0.50, variable), Spica (+0.96), Fomalhaut (+1.14), Deneb (+1.26), and Regulus (+1.35). MacPherson (1881:72-75) speculates that the reason for this is found in a Boorong preference to systematically group stars. He proposes that characters represented by stars of a particular family are either:

1) Grouped based upon their arrangement in the sky, specifically grouping three stars (or clusters) in a linear pattern;
2) Grouped into four linear arrangements that are roughly parallel to each other; or
3) Arranged roughly parallel to the horizon as they rise in the evening sky in their respective seasons at the latitude of the region (36° S).

**Table 1:** Stars in Stanbridge (1858, 1861) not identified by name, which we identify using his descriptions.

| Boorong Name | Stanbridge's Description | Name | Bayer designation |
|---|---|---|---|
| Karik Karik | the two stars in the end of the tail of Scorpio | Lesath<br>Shaula | υ Scorpii<br>λ Scorpii |
| Bermbermgle | the two large stars in the forelegs of Centaurus | Rigil Kent<br>Hadar | α Centauri<br>β Centauri |
| Tchingal | the dark space between the forelegs of Centaurus and Crux | Coalsack … | |
| Bunya | star in the head of Crux | Gacrux | γ Crucis |
| Kulkunbulla | stars in the Belt and Scabbard of Orion | Alnitak<br>Alnilam<br>Mintaka<br>…<br>…<br>Orion Nebula | ζ Orionis<br>ε Orionis<br>δ Orionis<br>ι Orionis<br>θ$^{1,2}$ Orionis<br>… |
| Weetkurrk | star in Boötes, west of Arcturus | Muphrid | η Boötis |
| Collowgullouric War | large red star in Rober Carol, … marked 966 | | η Carinae† |
| Collenbitchick | Double Star in the head of Capricornus | Prima Giedi<br>Secunda Giedi | α$^1$ Capricorni<br>α$^2$ Capricorni |
| Unurgunite | Star marked 5th mag 22 between two larger ones in the body of Canis Major | … | σ Canis Majoris |
| Wives of Unurgunite | The stars on either side of Unurgunite are his two wives | Adhara<br>Wezen | ε Canis Majoris<br>δ Canis Majoris |
| Wives of Totyarguil | Two stars on either side of Aquilla [Altair] | Tarazed<br>Alshain | γ Aquilae<br>β Aquilae |

† Also known as η Roburis or η Argus

Examples of this linear grouping include Orion's Belt, Aldebaran, and the Pleiades (Group 1), Vega, Altair, and α Capricorni (Group 2), Antares, Arcturus, Shaula, and Lesath (Group 3). In these cases (except for Shaula and Lesath), the characters represented by these stars are of a particular family. MacPherson asserts that because Procyon, Betelgeuse, Spica, Fomalhaut, Deneb, and Regulus do not fall into this systematic mechanical grouping, despite their brightness, the Boorong did not include them. We refer the interested reader to MacPherson (1881) for a more detailed, in-depth explanation of this systematic grouping of stars. We must also remember that Stanbridge's source was apparently two members of a single Boorong family (Stanbridge, 1861:301), and particular bright stars may have been omitted by either Stanbridge, or by his Boorong informants for unknown reasons.

Massola (1968) published "Bunjil's Cave", a book highlighting the oral traditions of various Aboriginal groups of Victoria, including the Wotjobaluk, Mara, Kulin, Kurnai, Bidwel, Ya-itma-thang, and Murray River communities (but does not specify Boorong oral traditions). The Wotjobaluk and Boorong are both clans of the Wergaia language. He claims to have obtained his information from fieldwork over a period of 10 years, as well as "the scant published material" (Massola, 1968:x**)**. Massola (1968:105) states that "beliefs of the Victorian Aborigines regarding the world appear to have been much the same amongst all tribes". On page 109 he cites stars in Wotjobaluk and Mara traditions not included in those of the Boorong, specifically Fomalhaut, which he describes as an eaglehawk ancestor, to which unspecified Murray River communities attribute to Mars.  As we discuss in Section 6, some of Massola's descriptions of Wotjobaluk sky knowledge appear to have been adopted directly from Stanbridge's paper.  Additionally, the Moporr (Mara) people included Betelgeuse in their sky knowledge, as opposed to the Boorong or Wotjobaluk (Dawson, 1881:100-101).  Betelgeuse is included on the oral traditions of other Aboriginal groups across Australia (see Norris & Hamacher, 2009; Maegraith, 1932), so it is unclear if MacPherson's hypothesis applies only to Wergaia oral traditions or if the Boorong simply did not tell Stanbridge about certain stars for whatever reason.  In Table 3 we present a comparison of Boorong, Wotjobaluk, Mara/Moporr, and Kulin star names.

**Table 2:** Summary table showing all stars and celestial objects specifically identified by Stanbridge, in both Western and Boorong nomenclature.  Stars are ordered by magnitude, from brightest to faintest.

| Western | Boorong | *V* mag | Constellation |
|---------|---------|---------|---------------|
| Sirius | Warepil | −1.46 | Canis Major |
| Canopus | War | −0.72 | Carina |
| η Carinae | Collowgullouric War | −0.4 † | Carina |
| Rigil Kent | Berm-berm-gle | −0.28 | Centaurus |
| Arcturus | Marpeankurrk | −0.03 | Boötes |
| Vega (Lyra) | Neilloan | +0.03 | Lyra |
| Capella | Purra | +0.08 | Auriga |
| Rigel | Collowgullouric Warepil | +0.15 | Orion |
| Achernar | Yerrerdetkurrk | +0.45 | Eridanus |
| Hadar | Berm-berm-gle | +0.61 | Centaurus |
| Acrux | Tchingal [spear in neck] | +0.75 | Crux |
| Altair | Totyarguil | +0.76 | Aquila |
| Aldebaran | Gellarlec | +0.86 | Taurus |
| Antares | Djuit | +0.98 | Scorpius |
| Pollux | Wanjel | +1.16 | Gemini |
| Mimosa | Tchingal [spear in rump] | +1.25 | Crux |
| Adhara | *Wife of Unurgnnite* | +1.50 | Canis Major |
| Castor | Yurree | +1.58 | Gemini |
| Gacrux | Bunya | +1.62 | Crux |
| Shaula | Karik Karik | +1.64 | Scorpius |
| Alnilam | Kulkunbulla | +1.70 | Orion |
| Alnitak | Kulkunbulla | +1.74 | Orion |
| Wezen | *Wife of Unurgunite* | +1.82 | Canis Major |
| Mintaka | Kulkunbulla | +2.23 | Orion |
| Muphrid | Weetkurrk | +2.65 | Boötes |
| Lesath | Karik Karik | +2.69 | Scorpius |
| γ Aquilae | *Wife of Totyarguil* | +2.72 | Aquila |
| ι Orionis | Kulkunbulla | +2.77 | Orion |
| τ Scorpii | *Wife of Djuit* | +2.82 | Scorpius |
| σ Scorpii | *Wife of Djuit* | +2.90 | Scorpius |
| σ Canis Majoris | Unurgunite | +3.47 | Canis Major |
| α² Capricorni | Collenbitchick | +3.56 | Capricornus |
| β Aquilae | *Wife of Totyarguil* | +3.71 | Aquila |
| θ$^{1,2}$ Orionis | Kulkunbulla | +3.9†† | Orion |
| α¹ Capricorni | Collenbitchick | +4.24 | Capricornus |

† Approximate magnitude in 1850.       †† Combined magnitude of individual components

**Table 2:** *continued…*

| Western | Boorong | Western | Boorong |
|---------|---------|---------|---------|
| *Solar System* | | *General* | |
| Sun | Gnowee | Space | Tyrille |
| Moon | Mityan | Star | Tourte |
| Jupiter | Ginabongbearp | Milky Way | Warring |
| Venus | Chargee Gnowee | Magellanic Clouds | Kourtchin |
| Meteor | Porkelongtoute | Coal Sack | Tchingal |

| Constellations or Groups | | Seasons | |
|---|---|---|---|
| Delphinus | Otehocut | Autumn | Weeit |
| Coma Berenices[Ψ] | Tourtchinboionggerra | Summer | Cotchi |
| Pleiades | Larnankurrk | Winter | Myer |
| Corona Australis | Won | Spring | Gnallew |
| [Ψ] Given as Cornua Berenices in Stanbridge (1858:139) or Coma Berenices in Stanbridge (1861:302) | | | |

Several studies (e.g. Fredrick, 2008; Johnson, 1998) have shown that many Aboriginal groups gave significance to the brightest individual stars, their nearby companion stars, naked-eye double stars, small distinctive asterisms or clusters, and the dark dust clouds silhouetted along the Southern Milky Way. Several Aboriginal groups noted compact, but distinctive groups of relatively faint stars: for example, Mountford (1956:479) and Haynes (1992, 2000:58-59) have related how the Aboriginal people of Groote Eylandt gave the name of Unwala (the Crab) to the compact group of third and fourth magnitude stars that comprise the head of Hydra, midway between the first-magnitude stars Procyon and Regulus, the latter two stars apparently not considered significant. In contrast, Peter Beveridge, reporting to the Select Committee of the Legislative Council of Victoria, stated that "[the Aboriginal Victorians] have a name and legend for every planet and constellation visible in the heavens" (Ridley, 1873b: 278), although this may have been a generalised statement that was not deeply researched by Beveridge.

These same general trends are also seen in the oral traditions of the Boorong, who identified all but five of the 21 first-magnitude stars, and a total of over 30 individual stars, plus Delphinus, Coma Berenices, the Magellanic Clouds, Venus, and Jupiter. There remains some confusion in the literature over the identity of the Boorong object called "Won", identified simply as "Corona" by Stanbridge (1858), representing the boomerang thrown by Totyarguil (Altair). We note that Corona Australis has nearly the same right ascension as Altair and is relatively near to it on the sky, so we identify "Won" as Corona Australis, rather than Corona Borealis (cf. Massola 1968; Johnson 1998). Of the two constellations, Corona Australis has the more geometrically symmetric pattern, and is rather like a Boomerang, representing the only apparent instance where a 'connect-the-dots' star pattern is applied by the Boorong (see Footnote 1). Corona Australis is quite distinctive under a dark sky, despite the relative faintness of its stars.

**Table 3:** Comparative names of celestial bodies taken from Stanbridge (1861) for Boorong and Massola (1968) for the Wotjobaluk, Mara, and Kulin groups of Victoria. All objects are listed alphabetically.

| Object | Boorong | Wotjobaluk | Mara/Moporr | Kulin |
|---|---|---|---|---|
| *Stars* | | | | |
| Achernar | Yerrer-det-kurrk | … | … | … |
| Aldebaran | Gellarlec | Gallerlec | … | … |
| β Aquilae | Wife of Totyarguil | Wife of Totyarguil | … | Kunnawarra |
| γ Aquilae | Wife of Totyarguil | Wife of Totyarguil | … | Kunnawarra |
| Altair | Totyarguil | Totyerguil | … | Bunjil |
| Antares | Djuit | Djuit | Butt Kuee tuukuung † | Balayang |
| Arcturus | Marpeankurrk | Marpean-kurrk | … | … |
| Betelgeuse | … | … | Moroitch † | … |
| α[1,2] Capricorni | Collen-bitchik | Collenbitchik | … | … |
| α Centauri | … | Purt-mayel | … | Djurt-djurt |
| β Centauri | … | Bram-bram | … | Thara |
| σ Canis Majoris | Unurgunite | Urnugunite | … | … |
| Canopus | War | War | Waa † | Lo-an-tuka |
| Capella | Purra | Purra | … | … |
| η Carinae | Collowgulloric War | Collow-collouricwar | … | … |
| Castor | Yurree | Yurree | … | … |
| δ Crucis | … | Dok †† | … | … |
| γ Crucis | Bunya | Bunya | … | … |
| Fomalhaut | … | [recorded, no name] | Buunjill † | … |
| Muphrid | Weetkurrk | Weet-kurrk | … | … |
| Pollux | Wanjel | Wanjel | … | … |
| Rigel | Collowgulloric Warepil | Yerrerdet-kurrk | … | … |
| 2 stars in Sag. | … | … | … | Tadjeri |
| | … | … | … | Tarnung |
| λ, υ Scorpii | Karik Karik | Karik Karik | Kummim bieetch † | … |
| Sirius | Warepil | Warepil | Gneeanggar † | Lo-an |
| Tarazed | … | Wives of Totyarguil | … | Kunnawarra |
| Vega | Neilloan | Neil-loan | … | … |
| Unknown Yellow Star in Orion | … | … | Kuupartakil † | … |
| *Clusters & Star Groups* | | | | |
| Beehive Cluster | … | Coomartoorung | … | … |
| Orion's Belt | Kulkunbulla | Kulkunbulla | Kuppiheear † | … |
| Pleiades | Larnankurrk | Larnan-kurrk | Kuurokeheear † | Karatgurk |
| Pointers | Berm-berm-gle | Bram-bram-bult | Tuulirmp † | … |
| *Constellations* | | | | |
| Coma Berenices | Tourtchinboionggerra | Tourt-chimboion-gherra | … | … |
| Corona Australis | Won | Wom | … | … |
| Crux | … | … | Torong (or) | … |
| | … | … | Kunkun Tuuromballank † | … |
| Delphinus | Otchocut | Otchout | … | … |
| Hydra (head?) | … | … | Barrukill † | … |
| *Galaxies & Nebulae* | | | | |
| Coalsack | Tchingal | Tchingal | Torong † | … |
| LMC | Kourt-chin | [recorded, male] | Kuum Kuuronn † | … |
| Milky Way | Warring | [smoke, name not given] | Barnk † | … |
| SMC | Kourt-chin | [recorded, female] | Gnaerang Kuuronn † | … |
| *Solar System* | | | | |
| Comet | … | … | Puurt Kuurnuuk † | … |
| Jupiter | Ginabongbearp | Ginabonbearp | Burtit Tuung Tirng † | … |
| Mars | … | … | Parrupum † | … |
| Meteor | Porkelongtoute | … | Gnummae waar † | … |
| Moon | Mityan | Mityan | Meeheaarong Kuurtaruung † | Menyan |
| Sun | Gnowee | Gnowee | Tirng † | … |
| Venus | Chargee-gnowee | Chargee-gnowee | Wang'uul/Paapee Neowee † | … |

† Moporr (Mara) names taken from Dawson (1881).
†† Massola (1968:8) refers to Dok (the frog) as the mother of the Bram-bram-bult (the Pointers), represented by a star in Crux closest to the Pointers, referring to β Crucis. However, on page 108, he claims Duk [*sic*] is the west star of Crux, referring to δ Crucis, and that α and β Crucis are represented by spears thrown by the Bram-bram-bult that pierced the emu (Tchingal – Coalsack). Both accounts are apparently taken from the Wotjobaluk clan.

## 4    Identification of η Carinae

In Stanbridge (1858:140), he describes a bright star called *Collowgullouric War*, as a female crow, the wife of War (Canopus). He labels it as "a large red star in Rober Carol [*sic*]", and gives the identification number "966"; an extract is reproduced here as Figure 3. We deduce that Collowgullouric War is referring to η Carinae in outburst, during the period of Stanbridge's fieldwork, c. 1848 to 1851, which coincides with the years during which η Carinae was at its brightest (Smith & Frew, 2010), and that this outburst was incorporated into Boorong oral traditions. We expand on the reasoning underpinning our identification below.

140            *On the Astronomy of the Aborigines.*

Won (Corona), a boomerang thrown by Totyarguil.
Weetkurrk (Star in Bootes, west of Arcturus), daughter of Marpeankurrk.
War (Male Crow) (Canopus), the brother of Warepil, and the first to bring down fire from (tyrille) space, and give it to the aborigines, before which they were without fire.
Collowgullouric War (large red star in Rober Carol, marked 966) (Female Crow), wife of War. All the small stars around her are her children.

**Figure 3:** An excerpt from Stanbridge (1858:140) describing War (Canopus) and his wife, Collowgullouric War.

"Rober Carol" refers to the now-defunct constellation of *Robur Carolinum* (Latin for "Charles' Oak") created by Sir Edmond Halley[3] in 1679, after observing the southern sky from St Helena in 1677 (Halley 1679; Baily, 1843). This new constellation was appropriated from the classical star group of Argo Navis (also now defunct), and constituted stars now located in eastern Carina and Vela, and western Centaurus (see Figure 4). Halley (1679) recorded η Carinae as a fourth magnitude star (see Figure 5), and Frew (2004) showed its magnitude at that time was $V = +3.3 \pm 0.3$ on a modern photometric scale.

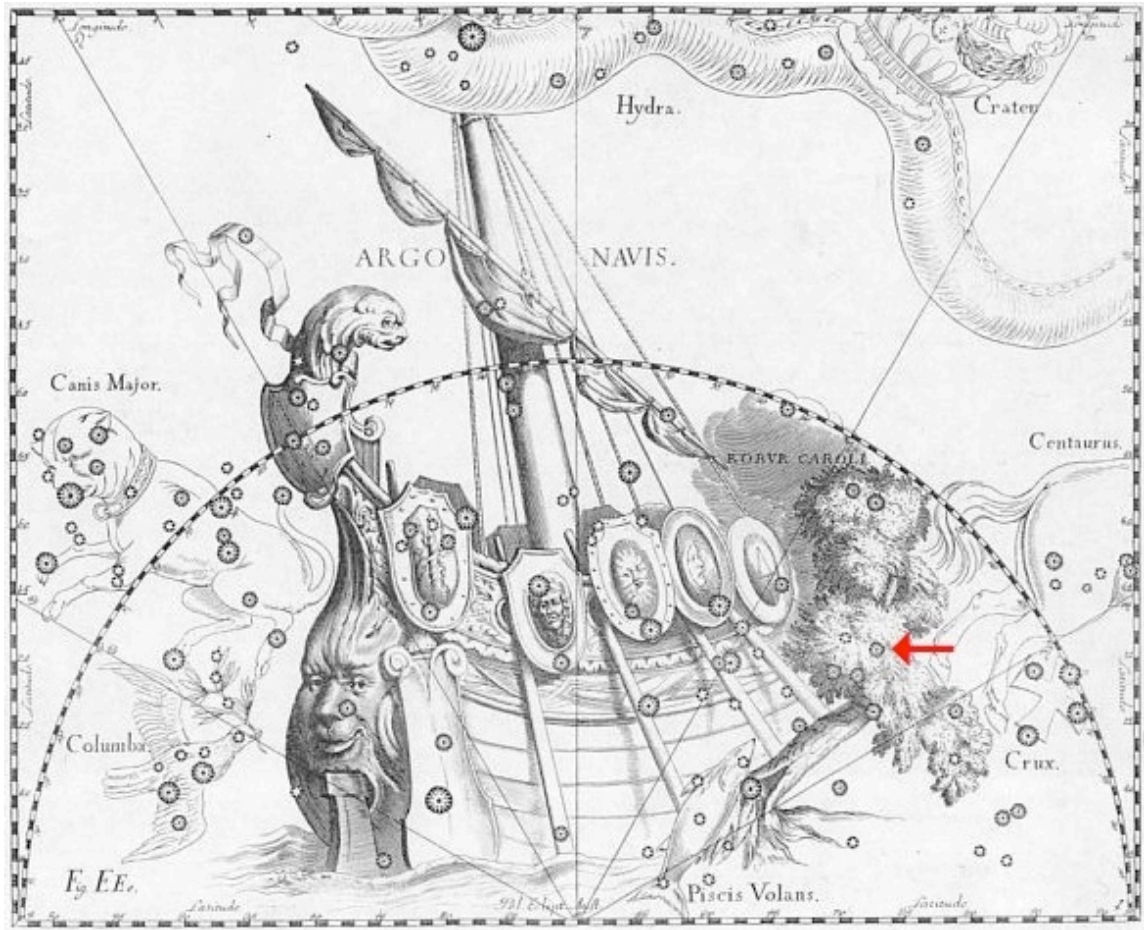

**Figure 4:** A star chart featuring *Robur Carolinum* from the star atlas of Jan Hevelius, who utilized the data from Halley (1679). The image is adapted from the modern facsimile of this atlas by Sheglov (1968:53). Eta Carinae has been marked with a red arrow.

Stanbridge cites the simple designation '966'. This certainly refers to a designation from La Caille's pioneering catalogue of 1,942 southern stars (La Caille, 1763). The reader is referred to Frew (2004:9) for the cross-identifications of stars in this region of the sky, taken from old catalogues. We further note that the correct designation for η Carinae from La Caille's catalogue is '968 Argus', but the surrounding Carina Nebula (NGC 3372) received the designation '966 Argus', even though it is a non-stellar, extended object. This small discrepancy in the designation is probably the result of a transcription error by Stanbridge (see Figure 6), but we can deduce from this designation alone that Stanbridge was referring to a bright red star associated with the Carina nebula, i.e. η Carinae itself.

**Figure 5:** An excerpt from Halley's (1679) Catalogue of Southern Stars, taken from the more accessible edition of Baily (1843:173). Entry 146 (Sequens) refers to the coordinates of Eta Car, with the designation '968' taken from Lacaille (1763). Halley noted η Carinae as a fourth magnitude star, which was shown by Frew (2004) to be equal to $V = 3.3 \pm 0.3$ on the modern scale.

**Figure 6:** Excerpt from La Caille (1763) showing the entry for the Carina Nebula (966), described as "*ne.*" or nebulous, and η Carinae (968), listed as a second magnitude star with the designation η. We highlight these entries with a red arrow.

However, it would seem unlikely that Stanbridge had a copy of La Caille's (1763) catalogue, which is a very rare work. η Carinae was not cross-referenced with La Caille (1763) in the British Association Catalogue, but was instead listed as 'Lac 4457' from the complete edition of La Caille's catalogue of 9,766 stars. However, the Paramatta or 'Brisbane' catalogue does cross reference η Argus as Lac 968, listing it as a second magnitude star, calling it η Argus. Perhaps Stanbridge had access to Johann Bode's *Uranographia* atlas (Bode, 1801), which shows the outline of Robur Carolinum, marked as Robor Caroli. In summary, it is unclear exactly what atlas and catalogue Stanbridge used to make his identifications. Nonetheless, this lack of knowledge does not affect our conclusions.

In the two decades prior to the publication of Stanbridge's paper, η Carinae was one of the brightest stars in the sky and would have demanded attention from even a casual skywatcher (Frew, 2004). Contemporary accounts commonly referred to its colour as orange or reddish during the Great Eruption (e.g. Smyth, 1845; Jacob, 1847; Gilliss, 1855, 1856; Moesta, 1856; Abbott, 1861; Powell, 1862; Tebbutt, 1866; see Smith &

Frew, 2010), consistent with Stanbridge's depiction of it as "a large red star". A cursory view of this region of sky with the unaided eye shows this to be one of the richest regions of stars in the southern Milky Way (called *Warring* in the Boorong language). Several third to fifth magnitude stars are located within a few degrees of η Carinae and are very likely the "small stars" (children) referred to in Stanbridge's account.

There is only one other star in Robur Carolinum that is reddish or orange in hue and is brighter than the fourth magnitude, which is q Carinae ($V = +3.32$, spectral type K3Ib). Of the 34 stars positively identified in Stanbridge's study (see Table 2), only three were below third magnitude ($V > +3.5$), of which two are components of a conspicuous naked eye double star ($\alpha^1$ and $\alpha^2$ Capricorni). It is unlikely that q Carinae is the star recorded by Stanbridge, but is more likely to be one of the "children". Furthermore, this star is rather too faint for its colour to be *obvious* in unaided vision. While the naked eye limit for human foveal (direct) vision is $V \approx +4.1$ (Schaefer, 1993) or perhaps a little fainter (Schaefer, 1996), stars much fainter than $V \approx +3.0$ fall in the domain of mesopic vision, where colour perception is quite poor (Malin & Frew, 1995:86). This is further evidence that the "large red star", recorded by Stanbridge (and by extension the Boorong) was η Carinae, and not q Carinae or some other fainter star.

Given the brightness and colour of η Carinae during the years of Stanbridge's fieldwork, its location in Robur Carolinum, and considering its designation in La Caille's (1763) catalogue and those catalogues that cross-references La Caille, we determine that Collowgullouric War is a Boorong record of η Carinae during its period of outburst in the 1840s.

## 5      About η Carinae

η Carinae is a luminous hypergiant at a distance of 2350 ± 50 pc from the Sun (Smith, 2006) in the constellation Carina (J2000, α: $10^h\ 45^m\ 04.0^s$, δ: −59° 41′ 04″). It is a massive binary system with a combined mass exceeding 100 times that of the Sun. The current visual magnitude is $V \approx +4.6$ (Fernández-Lajús et al., 2009; Verveer & Frew, 2009). The dominant member of the binary system is an eruptive luminous blue variable star, with a luminosity of approximately four million times that of the sun (Davidson & Humphreys, 1997).

η Carinae has varied markedly since its first recorded observation over four centuries ago (Figure 7, taken from Frew, 2004). It was a first or second magnitude star at the beginning of the nineteenth century, before John Herschel made the first detailed series of brightness measurements in the 1830s (Herschel, 1847). He noted that its brightness was relatively constant during this period ($V = +1.2$ on the modern Pogson scale), before he observed it to rapidly brighten at the close of 1837 to be as bright as Alpha Centauri, before quickly fading again — this brightening is generally considered to be the start of the period of enhanced brightness known as the Great Eruption (Frew 2004; Smith & Frew, 2010)[4]. η Carinae brightened markedly again during 1843; at its peak brightness in March of that year, and again in January 1845, it was the second brightest star in the sky after Sirius. The well-known Homunculus nebula is the ejected debris from this explosive event (e.g. Thackeray, 1949; Gaviola, 1950; Smith & Gehrz, 1998; Walborn et al., 1978; see Figure 8). The origin of this eruption, sometimes called a "supernova impostor" event, remains uncertain (e.g. Smith, 2008; Smith and Owocki, 2007). The star returned to its pre-outburst brightness in 1858, and continued to rapidly fade; by 1869, the star was invisible to the naked eye.

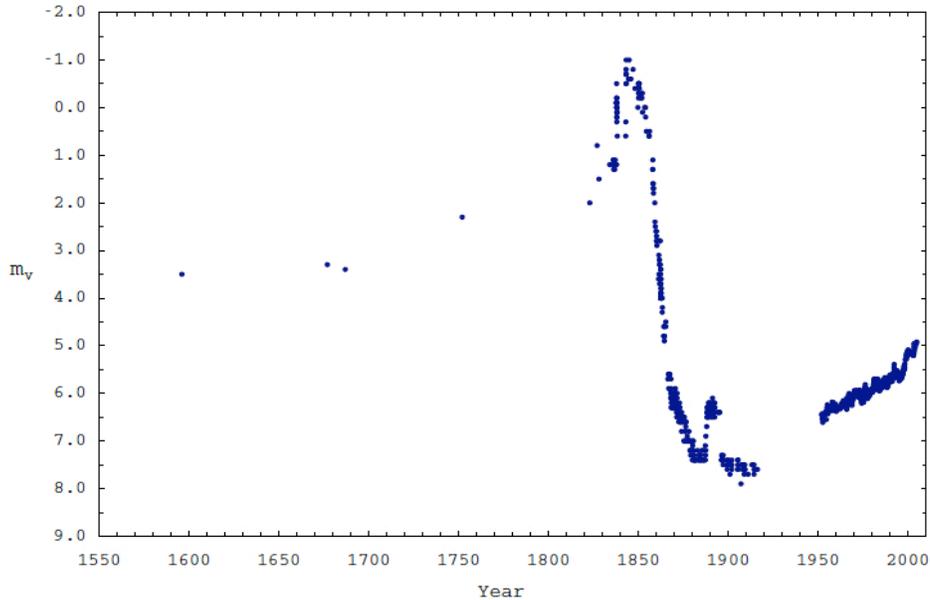

**Figure 7:** The visual light curve of η Carinae between 1596 and 2000 (from Frew, 2004).

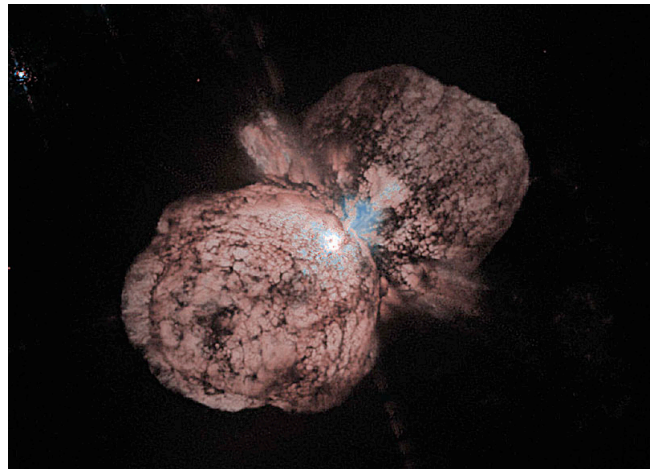

**Figure 8:** A NASA *Hubble Space Telescope* WFPC2 image of η Carinae (at centre) and its expanding, bipolar Homunculus nebula ejected during the Great Eruption in the 1840s, image is ~25 arcsec in width (Credit: J. Morse & K. Davidson, 10 June 1996).

## 6        Previous Attempts to Identify Collowgullouric War

MacPherson (1881:73) explains that the female crow is the "small red star No. 966 in King Charles' Oak [Robur Carol]". It is interesting that he would describe this star as "small", considering Stanbridge had specifically referred to it as "large". However, by the 1880s, η Carinae was much fainter with an apparent magnitude of *V* = +7.4 (see Figure 9), too faint to be discerned with the naked eye against the backdrop of the rich Carina nebula. In addition, Johnson (1998:122), who cited MacPherson, described Collowgullouric War as "a small red star, probably Epsilon (ε) Carinae" (*V* = +1.86), to which Haynes (2000:75) also agrees. It is possible that MacPherson attributed Collowgullouric War to a "small" red star because of the faintness of η Carinae during the time he published his paper (see Figure 9). Both Johnson and Haynes may have been unaware of η Carinae's variable past, and instead identified Collowgullouric War as ε Carinae because of its current brightness and slightly orange tint.

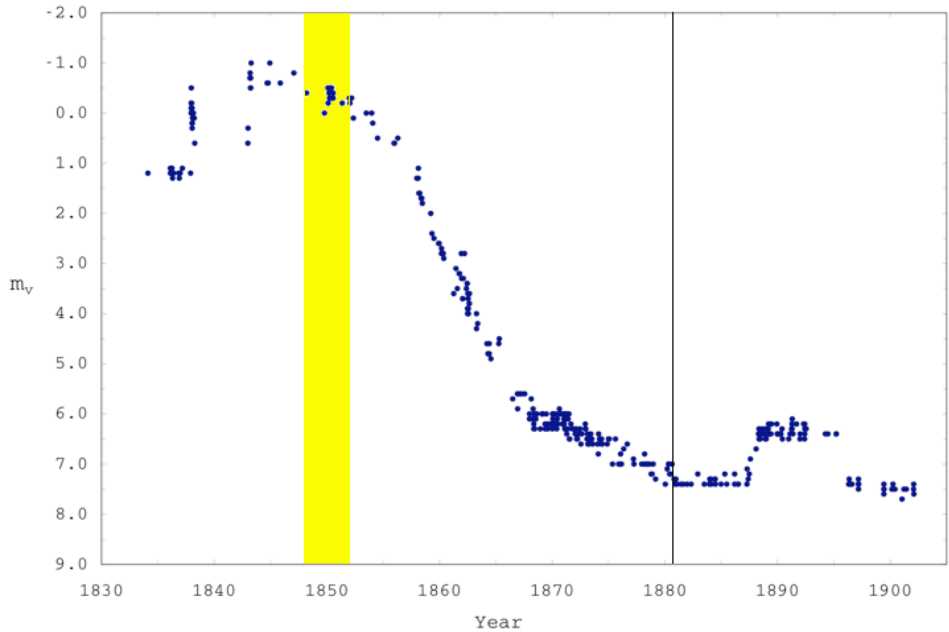

**Figure 9:** The visual light curve between 1830 and 1900, covering the period of the Great Eruption that extended from c. 1837 to 1857 (from Frew, 2004). The time period in which Stanbridge was likely to have conducted his fieldwork (approximately 1847 to 1851) is denoted by the vertical bar. The year in which MacPherson described η Carinae as a "small red star" (1881) is highlighted by the vertical line.

Stanbridge (1861:303) said that neighboring Aboriginal groups, from Swan Hill (near Tyrell Downs) to Mount Franklin (near Daylesford), share similar names and associations of the stars with the Boorong. A perusal of Massola (1968) shows that the Wotjobaluk clan of the Wergaia language group, of which the Boorong is also a clan, has nearly identical views of the night sky as the Boorong. Although Stanbridge does not provide the details of stories associated with Boorong stars, Massola does concerning Wotjobaluk stars (1968:3-27). Massola identifies the crow (War) in Wotjobaluk traditions as Canopus and mentions that the largest star in "Rober Carol" is "Collow-collouricwar", his wife (1968:109); a reference to what we identify as η Carinae. In this case it is likely that Massola referred directly to Stanbridge's accounts, as most of the Wotjobaluk names as given by Massola are almost identical to the Boorong names recorded by Stanbridge. Furthermore, some descriptions are identical (e.g. Muphrid or Weetkurrk is described as a "star in Boötes, west of Arcturus" by both Stanbridge and Massola). Interestingly, Massola noted some celestial objects not mentioned by Stanbridge that have significance to the Wotjobaluk, such as Fomalhaut and the Beehive cluster (a conspicuous naked-eye star cluster, also called M44 or NGC 2632), although the reason for their exclusion in Boorong astronomy is unclear. It is worth mentioning that although Massola (1968:109) mentions Collow-collouricwar as the wife of War, he does not include any mention of Collowgullouric War in the Wotjobaluk stories (Massola, 1968:3-27).

In his unpublished MA thesis, Morieson (1996:74) identified Collowgullouric War as η Carinae, but did not reveal his reasoning for this. However, he claims that other nearby red stars, including R Carinae, S Carinae, and a fifth magnitude star near the open cluster NGC 2516, may also be Collowgullouric War (but excludes this possibility in later publications). We can rule these out as candidates, since they do not match Stanbridge's description as "large" (bright). Furthermore, R and S Carinae are both large-amplitude, Mira-type variable stars, spending most of their time below naked-eye visibility. And as mentioned earlier, stars of these magnitudes do not show apparent colour to the unaided eye. Finally, R Carinae, S Carinae, and NGC 2516 are not labeled "966" in any star catalogues, leaving no alternative star to identify as Collowgullouric War.

In summary, we consider the evidence for a Boorong identification of η Carinae during its Great Eruption to be unambiguous, as there are no alternative bright red stars nearby in the sky from which to choose, nor any others labeled 966.

## 7    Incorporation of η Carinae into Boorong Sky Knowledge

We further argue that η Carinae was probably not in the oral traditions of the Boorong prior to its outburst, and propose that this transient "supernova impostor" event was significant enough to be included in their oral traditions during the 1830s or early 1840s. In Stanbridge's account, the Boorong associate Warepil, Collowgullouric Warepil, and War with some of the brightest stars in the sky (namely Sirius: $V = -1.46$, Rigel: $V = +0.15$, and Canopus: $V = -0.72$, respectively). This suggests that they chose η Carinae to represent Collowgullouric War given it was of similar magnitude during its Great Eruption ($0 > V > -1.0$).

Prior to its major outburst (i.e. before 1820), η Carinae was a second or third magnitude star in a region of the sky densely populated with stars of similar or brighter magnitudes, such as β Carinae ($V = +1.68$), ε Carinae ($V = +1.86$), δ Velorum ($V = +1.94$), ι Carinae ($V = +2.25$) and κ Velorum ($V = +2.50$). We emphasise that Stanbridge recorded none of these conspicuous stars.

An interesting trend regarding the magnitude of particular stars and their celestial spouses is apparent in Boorong traditions. Masculine stars with a single wife are both represented by objects of comparable brightness and are always in the same region of the night sky (see Table 4). If the husband has two wives, they are of different magnitudes, usually being fainter than the husband, except in the case of Unurgunite where the wives are brighter. Additionally, if a character has two spouses, the trio is found to be in a fairly straight line (see Djuit, Totyarguil, and Unurgunite in Table 4). Such a linear preference in grouping stars was described by MacPherson (1881:73-75), as discussed before.

**Table 4:** Comparative brightness between celestial objects representing husbands and those representing their wife or wives.

| Husband | Object | V-mag | Spouse(s) | V-mag |
|---|---|---|---|---|
| Warepil | Sirius | −1.46 | Rigel | +0.15 |
| Ginabongbearp | Jupiter | −2.9 † | Venus | −4.8 † |
| War | Canopus | −0.72 | η Carinae | −1.0 † |
| Kourt-chin | LMC | +0.4 | SMC | +2.3 |
| Djuit | Antares | +0.98 | τ Scorpii | +2.82 |
|  |  |  | σ Scorpii | +2.90 |
| Totyarguil | Altair | +0.76 | β Aquilae | +3.71 |
|  |  |  | γ Aquilae | +2.72 |
| Unurgunite | σ Canis Majoris †† | +3.47 | ε Canis Majoris | +1.50 |
|  |  |  | δ Canis Majoris | +1.82 |

† At peak brightness
†† This star is also called "22 Canis Majoris", identified as such by Stanbridge.

The bright southern portion of the Milky Way in the vicinity of η Carinae is especially rich in moderately bright naked eye stars, but the only first magnitude stars in that region belong to Crux. Considering the magnitude of Canopus (War), we would expect to find his wife represented by a star of similar magnitude relatively nearby in the sky. We know that during its nineteenth century outburst, η Carinae was comparable in brightness with Canopus and is in the same general region of the sky. Therefore, it seems it was deliberately chosen to be War's wife, with its brightness as a key factor. There is no obvious reason why η Carinae before its outburst would have been incorporated into celestial oral traditions so closely associated with the brightest stars in the night sky. Prior to the early 1800s, η Carinae was a second or third magnitude star (see Figure 9) in a region of the sky full of stars of similar magnitudes. It was only during the 20-year period between 1837 and 1857 that it became one of the brightest stars in the night sky before fading from view by the 1870s.

## 8    Incorporation of Celestial Phenomena into Oral Traditions

It is not uncommon for special events to serve as the foundation for new oral traditions or to be incorporated into pre-existing oral traditions (see Ross, 1986 for an in depth study of Australian oral traditions). Aboriginal oral traditions are not static, but rather dynamic and evolving. While Aboriginal oral traditions in most cases serve to illustrate and record a particular moral charter and to preserve the laws, cosmology, and social structure of the community, specific mnemonics can change over time. Several examples of transient celestial phenomena, including comets, meteors and meteorite falls, being incorporated into pre-existing oral traditions or serving as the foundation of others can be found throughout the literature (e.g. Hamacher & Norris, 2009; 2010; 2011), a few of which we discuss below.

The fall of a meteorite near Jupiter Well, Western Australia was incorporated into a new oral tradition (Poirier, 2005:237-238), as was an apparent meteorite fall witnessed by Aboriginal people and described to Barker (1964:109-110). The Aboriginal informant's description provided enough details to leave little doubt in Barker's mind that they had witnessed the fall. But since the meteorite was used as a source for sacred stories, the informants would not reveal its location to Barker or the other white Australians (for more examples, see Hamacher & Norris, 2009, 2010). In the 1800s, the Western Arrernte people of Ntaria (Hermannsburg, NT) incorporated Christian mythology into their pre-existing oral traditions during a period of conversion by Lutheran missionaries (see Austin-Broos, 1994). Some of these traditions were related to celestial phenomena, including a story of a falling star and rock art that depicted the sun, moon and stars (*ibid*).

In some cases, the appearance of a comet coincided with a natural disaster or catastrophic event. For example, the appearance of bright comets in Australian skies coincided with droughts (Parker, 1905:99), disease epidemics (Spencer & Gillen, 1899:549), war (Morrill, 1864:61) or natural disasters (Mowaljarlai & Malnic, 1993:194), prompting Aboriginal people to view the phenomenon with fear and apprehension (which is consistent with many indigenous cultures around the world, see Hamacher & Norris, 2011). These views were carried through successive generations through oral tradition. Even solar eclipses, which are rare occurrences from any given place on earth, have been incorporated into oral traditions (Bates, 1944; Johnson, 1998; Norris & Hamacher, 2009; Warner, 1937). Therefore, the hypothesis that transient celestial events are incorporated into Aboriginal oral tradition is supported in the literature.

Without Stanbridge's description of Collowgullouric War's location and the catalogue designation he quoted, it would be difficult to argue that the Boorong were describing η Carinae. We have no records prior to Stanbridge's papers regarding Boorong astronomy, and indeed none since. The Boorong clan no longer exists as an independent entity, but their descendents still live in the region as members of the Kulin nations (Clark, 1990). It would be of interest to know how the oral traditions regarding these stars changed, if at all, as η Carinae faded to invisibility just a decade after Stanbridge's fieldwork.

# 9 Are there other indigenous records of η Carinae?

There are no unambiguous records of η Carinae from antiquity, and the earliest observations are derived from Dutch explorers at the close of the sixteenth century (Frew, 2004). The presence of faint nebulous ejecta exterior to the Homunculus nebula suggests that η Carinae underwent a putative earlier outburst that occurred several centuries before the Great Eruption (Walborn et al., 1978; Walborn & Blanco, 1988; Smith & Morse, 2004), perhaps around 1000 C.E.

We have shown that the Boorong recorded η Carinae during its outburst in the nineteenth century, when it became the second brightest star in the night sky. Indeed, the long duration ($\geq$ 20 yrs) of the Great Eruption[4], far longer than any supernova event, and its sheer brightness compared to neighbouring stars, suggest it would have been widely observed by most, if not all, indigenous peoples of the southern hemisphere. Hence, we have examined a number of accounts summarising the ethnoastronomy of Australian indigenous groups that post-date Stanbridge's work, but nevertheless we have found no clear evidence for any other accounts that refer to an unidentified red star in Carina (e.g. Beveridge, 1889; Dawson, 1881; Howitt, 1904; Maegraith, 1932; Massola, 1968; Manning, 1882; Mathews, 1904; Mountford, 1956; 1958; Palmer, 1884; 1886; Parker, 1905; Piddington, 1930; Ridley, 1873a; Stone, 1911; and Tindale, 1937, amongst other sources). Since the fieldwork for these studies dates from c. 1870 to 1950, the simplest explanation is that as the star faded below naked-eye visibility in the 1860s, any oral traditions based on η Carinae that were authored during the Great Eruption were possibly lost, or seen to be no longer important.

In addition, we ask if there are any archaeological records of the putative earlier outburst of η Carinae? Teames (2002) cites three stone artifacts from the pre-Incan Tiahuanacan culture (ca. 1000 C.E.) near the southern end of Lake Titicaca, Bolivia that may depict the 1000 C.E. outburst. While her evidence is highly intriguing, it is too open to interpretation to be definitive. Interestingly, Orchiston (2000; 2002) and Green & Orchiston (2004) have identified a reference in Best (1922) to a possible transient source in the southern Milky Way, recorded by the Maori of New Zealand or their Polynesian ancestors. The object is called *Mahutonga*, which Best (1922:46) records as "a star of the south that remains invisible". This reference in turn derives from Stowell (1911:202-203) who described Maahu (Mahu) as the star of the south, which "*has left its place in pursuit of a female. When it secures the female, it will come back again to its true home*". Stowell (1911:209) further states that "*Maahu-Tonga is invisible*". Based on this description, Green & Orchiston (2004) claim a transient event occurred in the region of *Mahu or Mahutonga*, Maori terms associated with the region of Crux and the Coalsack Nebula (Crux is the chamber of Maahu-Tonga; Stowell, 1911:209), and identify it as a potential supernova. While they were unable to identify a particular supernova event, they highlighted the possibility of it being a transient observed in 185 C.E., which is generally, but not universally, understood to be a supernova (Martocchia & Polcaro, 2009; Schaefer, 1995; Stephenson & Green, 2002; Zhao, Strom & Jiang, 2006). Its position (J2000, α: $14^h 43^m$, δ: −62° 27.7′), lies close to α Centauri near the border of Centaurus and Circinus.

However, the identification of Mahutonga with SN 185 would indicate that the account originated in Polynesia, over eight centuries before humans settled in New Zealand (~1000 C.E.). Stowell's description might be taken to describe a star that appears, disappears, and is expected to reappear again, suggesting a recurrent variable star as opposed to a supernova event that suddenly appears before fading on a timescale of months (Stephenson & Green, 2002). Given that η Carinae's brightness fluctuated significantly over several decades during the nineteenth century, it is possible that Mahutonga may be a reference to this. Since η Carinae is at a comparable angular separation to Crux as SN 185, η Carinae remains a possible, albeit speculative, candidate for *Mahutonga*.

## 10        Conclusion

Given that transient celestial phenomena can be incorporated into either pre-existing Aboriginal oral traditions or form the basis of new, we conclude that the Boorong people observed η Carinae in the nineteenth century, which we identify using Stanbridge's description of its position in Robur Carolinum, its colour and brightness, its designation (966 Lac, implying it is associated with the Carina nebula), and the relationship between stellar brightness and positions of characters in Boorong oral traditions. In other words, the nineteenth century outburst of η Carinae was recognised by the Boorong and incorporated into their oral traditions. This supports the assertion that Aboriginal sky knowledge is dynamic and evolving, and not static. The observations by the Boorong represent the first and only definitive indigenous record of the Great Eruption of Eta Carinae identified in the literature to date.

**Notes**

[1] Many of the characters observed by the Boorong, identified as bright stars by Stanbridge (1858), have been considered by Morieson (1996; 2002; 2006) to represent patterns of stars, which sometimes include very dim stars. Since Aboriginal groups in southeast Australia generally avoided a "connect-the-dots" approach to grouping constellations (see Fredrick, 2008; Johnson, 1998; Massola, 1968), instead attributing individual stars to specific characters in their oral traditions, these proposed constellation patterns seem unlikely to have been recorded by the Boorong.

[2] In addition, Fredrick (2008:58) proposes that a Wiilman Dreaming story (southwest Western Australia), recorded by Bates (1904-1912), describes the variability of Betelgeuse:

"*Orion was a hunter of women, who was kept away from the sisters of the Pleiades by*

*their older sister, thought to be the head of Taurus. Orion is described as wearing a feathered headdress, a string belt (the stars of Orion's Belt) and whitened tassel (Orion's Scabbard) and having a red-ochre body. The older sister throws out fire from her body and moves towards Orion, as she moves towards him she lifts her left foot and frightens him so much that the red magic of his arm and body becomes faint for a while. The magic comes back eventually and the sister asks for help from her family. All of the surrounding stars laugh at Orion and the red-back spider (Rigel) is ready to bite Orion,*" (Fredrick, 2008:58).

While the variability of Betelgeuse was first announced by Herschel (1840), the observed visual range of *V* = +0.2 to +1.2 (Goldberg, 1984) is large enough to be noticed by a regular sky-watcher. Indeed, Wilk (1999) has suggested that the variability of Betelgeuse was known in pre-Classical Greece. This will be the topic of a future paper.

[3] This constellation (Charles' Oak) was named in honor of King Charles II who is claimed to have hidden in an oak tree for a full day, during his defeat by Oliver Cromwell in the Battle of Worcester in 1651 (see Ridpath, 1989:147).

[4] However, η Carinae appeared as bright as a first magnitude star as early as July 1827, when the naturalist William Burchell observed it to be "as large as a Crucis" (Herschel 1847:35; Frew 2004:24).

**Acknowledgements**

The authors would like to acknowledge the descendents of the Boorong people, the Darug people (the traditional custodians of the land on which Macquarie University is situated), Ray Norris, John Morieson, and Wayne Orchiston. This research has made use of the NASA Astrophysics Data System (ADS), the SIMBAD database, operated at the Centre de Données astronomiques de Strasbourg (France), the Mitchell Library in Sydney, the Macquarie University Library, the National Library of Australia, and the Australian Institute for Aboriginal and Torres Strait Islander Studies. Hamacher was funded by the Macquarie University Research Excellence Scholarship (MQRES) within the Department of Indigenous Studies at Macquarie University.

**About the Authors**

Duane Hamacher is a Ph.D. candidate in the Department of Indigenous Studies at Macquarie University in Sydney, Australia. After graduating in physics from the University of Missouri and obtaining a Master's degree in astrophysics from the University of New South Wales, he was awarded a Research Excellence Scholarship to study Aboriginal Astronomy at Macquarie University.  Duane is also an astronomy educator at Sydney Observatory.

Dr. David Frew is a Research Fellow in the Department of Physics & Astronomy at Macquarie University. He works mainly on the evolution of planetary nebulae and symbiotic stars, but has wider interests in the history of astronomy, including the application of modern techniques to archival data of variable stars.